\documentclass[preprint,aps,nofootinbib]{revtex4}
\usepackage{graphicx}
\usepackage[dvips]{color}
\input epsf
\begin{document}

\newcommand{\be}{\begin{equation}}
\newcommand{\ee}{\end{equation}}
\def\bq{\begin{eqnarray}}
\def\eq{\end{eqnarray}}


\title{\bf On the Relativistic Formulation of Matter}
\author{Ram Gopal Vishwakarma\footnote{Electronic Address: vishwa@matematicas.reduaz.mx}}

\address{Academic Unit of Mathematics\\
 Autonomous University of Zacatecas\\
 C.P. 98068, Zacatecas, ZAC,
 Mexico}

\begin{abstract}
A critical analysis of the relativistic formulation of matter reveals some
surprising inconsistencies and paradoxes. Corrections are discovered which
lead to the long-sought-after equality of the gravitational and inertial
masses, which are otherwise different in general relativity.

Realizing the potentially great impact of the discovered corrections,
an overview of the situation is provided resulting from the
newly discovered crisis, amid the evidences defending the theory.



\end{abstract}
\maketitle

\section{Introduction}
Modern theories of gravitation describe
gravity not as a `force' in the usual
sense but as a manifestation of the curvature of spacetime. This
renders the theories essentially geometrical in character with non-Euclidean
geometries playing the central role. Einstein's
revolutionary discovery of the principle
of
general covariance then suggests that a relativistic theory of gravitation
should be formulated using the language of tensors. This implies that the matter, which is considered
as the source of the curvature, 
should also be represented invariably  by a tensor,  leading to the  
usual relativistic
hydrodynamic description of matter encoded in the famous energy-momentum-stress tensor (energy density-momentum-stress tensor, to be more precise) $T^{\mu\nu}$
($\mu,\nu=0,1,2,3$) which has become an integral
part of all the relativistic theories of gravitation including the candidate
theories of quantum gravity. The simplest example of $T^{\mu\nu}$ belongs to the case of a perfect fluid (absence of viscosities and heat flow). In a coordinate system $x^\alpha$, specifying the metric tensor $g_{\mu\nu}$, the $T^{\mu\nu}$ of the perfect fluid is given by
\be
T^{\mu \nu}=(\rho +p) u^\mu u^\nu-p g^{\mu\nu},\label{eq:emtensor}
\ee
where $\rho$, the energy density of the fluid and $p$, its isotropic pressure, are measured by an observer in a locally inertial reference frame which happens to be moving with the fluid at the instant of measurement. In the considered coordinates $x^\alpha$, the fluid as a whole has 4-velocity $u^\alpha=dx^\alpha/ds$. 
 {\it It is important to note that the term $\rho$, appearing in} (\ref{eq:emtensor}), {\it includes not only the rest mass of the individual particles of the fluid but also their kinetic energy, internal energy (for example, the energy of compression, energy of nuclear binding, etc.) and all other sources of mass-energy} \cite{MTW} (excluding the energy of the gravitational field).
 
Modelling matter by $T^{\mu\nu}$ has revolutionized the way we used to think about the source of gravity.
As mass density is the source of gravity in Newtonian theory, we expect
energy density to take over this role in the relativistic generalization of
Poisson's equation. To our surprise however, all the ten (independent)
components of $T^{\mu\nu}$ become contributing source of gravity, making a
revolutionary prediction that not only the energy density but pressure and
momentum too have gravitational effects. Though the momentum of
the matter
disappears from the scene once one uses a coordinate system comoving with the
matter, however,
it is not possible to avoid the pressure which becomes an integral part of
the relativistic description of matter.

It is an undeniable fact that the standards of $T^{\mu\nu}$, in terms of
beauty, consistency and mathematical
completeness, do not match the
vibrant geometrical side of the modern theories of gravitation, perhaps not enough attention has been
devoted to $T^{\mu\nu}$ as it deserves. Einstein himself conceded this fact
as he
wrote about the standards of $T^{\mu\nu}$ in the general theory of relativity (GR):
``{\it GR is similar to a building, one wing} (the geometry) {\it of which is
made of fine marble}, {\it but the other wing} (the matter) {\it of
which is built of low grade wood}". 
The mystery of the gravitational effects of the non-conventional sources (i.e., the components of $T^{\mu\nu}$ other than the energy density), is seldom addressed at a sufficiently foundational level. It has not been realized, for example, that the formulation of matter in terms of $T^{\mu\nu}$, modifies at the deepest level  its well-known properties, so that the intuitive Newtonian knowledge needs revision upon closer investigation. Let us consider, for example, the gravitational effects of pressure, which is the focal point of this paper.

Since pressure enters $T^{\mu\nu}$ as a scalar, just like the energy density, and since the dimensions of the pressure are those of the energy density, it implies that the pressure (somehow) {\it contains energy density}. (To what degree it contains energy density, depends on the equation of state.) 
This is an entirely new physics! Albeit counter-intuitive, this is perfectly consistent with equation (\ref{eq:emtensor}) wherein the pressure appears as being added to the energy density. Further, this new physics is corroborated by many examples wherein pressure
contributes to the energy density of the fluid. One may mention some cases from GR, for example.
One may mention the pressure contributing to the active gravitational mass density $(\rho+3p)/c^2$ in the
Raychaudhuri equation
\begin{equation}
\frac{3\ddot{\ell}}{\ell}=2(\omega^2-\sigma^2)+\dot{u}^\alpha_{;\alpha}-\frac{4\pi G}{c^2}
(\rho +3p),\label{eq:raychaudhuri}
\end{equation}
which is the relativistic analogue of Poisson's equation and determines the average
contraction/expansion of a self-gravitating fluid in GR. Here $\ell$ represents the volume behaviour of the fluid and the kinematical quantities  $\omega$, $\sigma$ and $\dot{u}^\alpha$ respectively measure rotation, shear and acceleration of the fluid flow.
As another example, one
can mention the pressure contributing to the passive gravitational mass
density $(\rho+p)/c^2$ of the fluid in the Tolman-Oppenheimer-Volkoff equation
\begin{equation}
\frac{dp}{dr}+(\rho +p) \frac{d}{dr}\ln(\vert g_{00}|)^{1/2}=0,\label{eq:TOV}
\end{equation}
which represents the relativistic analogue of the classical hydrostatic equilibrium of a
star. Here $r$ is the radial coordinate and $g_{00}$ is the time-time
component of the static spherically symmetric metric modelling the isotropic
material of the star. Let us note that the active and passive gravitational masses
are not equal in the modern theories of gravitation. The binding energy of the gravitational field is believed to be responsible for this. However, why the contributions from the gravitational energy to the different masses are not equal, has remained a mystery in these theories. Similar examples can be given from the alternative theories of gravity as well (one can see \cite{PhysRep}, for a review on the extended theories of gravity).

Notwithstanding strongly supported by various examples, the prediction that pressure contains energy, faces paradoxes and inconsistencies. There is already a longstanding Tolman paradox \cite{Ehlers} which can be described as follows.



\section{Tolman Paradox}

Tolman \cite{tolman} has derived a  formula for the total energy $E$, including gravitational energy, of a static system:
\[
E=mc^2=\int (\mathcal{T}_0^0-\mathcal{T}_1^1-\mathcal{T}_2^2-\mathcal{T}_3^3)~d^3x,
\]
where  $\mathcal{T}_\mu^\nu=T_\mu^\nu \sqrt{-g}$ is the tensor-density corresponding to the energy density-stress tensor. For the energy density-stress tensor (\ref{eq:emtensor}), this formula reduces to
\be
E=mc^2=\int (\rho +3p)\sqrt{|g_{00}|}~dV,\label{eq:rmass} 
\ee
which applies to a wide class of cases including the one in which the matter is confined to some limited region. Here $g_{00}$ is the time-time
component of the metric (in an asymptotically flat coordinate system representing the spacetime at a sufficient distance from the material system of interest lying in the neighbourhood of the origin of the coordinate system) and $dV$ is the proper spatial volume element of the fluid sphere. In modern terminology, $m$ is the ADM mass measured by a faraway inertial observer.

The Tolman paradox (related with the consequences of the gravitational effect of pressure) can be described as the following. The matter ($p=0$) at rest in a container exhibits the total mass $m$. However, converting the matter inside the container into disordered radiation ($p=\rho/3$) would double the total mass. Or alternatively, a conversion of gamma rays of total energy $E=mc^2$, enclosed in the container, into electron-positron pairs would lead to a total mass half\footnote{As energy and momentum are conserved when two photons collide to give rise to an electron-positron pair, so, if all the photons pairs do not meet this requirement, only those will contribute to the paradox, which meet the requirement. While their momentum will balance each others', their energy will be converted to the rest mass of the created electron-positron pairs. However, the term $3p$ appearing in the Tolman integral (\ref{eq:rmass}) would remain unaccounted, and hence would contribute to the paradox.}
 of $m$. In both cases, the conservation of $E$ is seriously violated.

The Tolman paradox has not received much attention, as the formula (\ref{eq:rmass}) involves gravitational energy (ascertained by the presence of $\sqrt{-g}$ in (\ref{eq:rmass})) which is notorious for violating its (local) conservation. Thus one can readily claim that the gravitational energy is responsible for the above-mentioned violation of the conservation of $E$ in the Tolman paradox (though it is apparent that the paradox results from the term $p$, and not from  $\sqrt{-g}$ in equation (\ref{eq:rmass})). In addition to the Tolman paradox, there have also been claims for the violation of energy in the cosmological context (see, for example, \cite{harrison}). However, like the Tolman paradox,  these claims have also not attracted much attention, as these usually take refuge, for a possible resolution, in the gravitational energy, which has been of an obscure nature and a controversial history.

So, is it possible to ascertain, beyond doubts, that there is really some fundamental problem in the relativistic formulation of  
matter, for example, given by the energy density-stress tensor (\ref{eq:emtensor})? If so, this must be done without implicating gravitational energy. Let us try to do this in the following.

\section{New Paradoxes} 

In the following, we focus our attention on the relativistic formulation of matter given by the energy density-stress tensor $T^{\mu\nu}$, {\it without considering any particular theory of gravitation}. After all, all the modern theories of gravitation use this tensor to represent the matter source. Usually, this tensor may have contributions from, or may be supplemented by, other sources as well, for example, the scalar fields (representing generalized `matter') or the geometric `matter' resulting from the modification of the  effective Lagrangian of the gravitational field by adding higher-order terms in the curvature invariants (as is done in the extended theories of gravitation \cite{PhysRep}). To be more precise, we shall derive our results from the conservation of $T^{\mu\nu}$, i.e., $T^{\mu\nu}_{ ~ ~ ;\nu}=0$ which is identically satisfied, through the Bianchi identities, in the simple cases of the metric theories of gravitation, for example, GR. There are however, other theories, for example, the Brans-Dicke theory, wherein this conservation does not follow from the divergence of the left hand side of the field equations. However, it is a standard practice to assume  $T^{\mu\nu}_{ ~ ~ ;\nu}=0$ (as an extra degree of freedom) in this type of theories.  Hence, the results we shall obtain in the following, apply to almost all the theories of gravitation. In the following we shall limit our analysis to the simplest case of $T^{\mu\nu}$ belonging to the perfect fluid, so that the physics does not get shrouded in the complexity of a more general $T^{\mu\nu}$. However, this compromise would not cost much since the perfect fluid represents/approximates a great many macroscopic physical systems, including the ordinary matter, the scalar fields, the dark energy candidates and perhaps the universe itself.

As the central point of our analysis in the following will be the energy density-stress tensor of a perfect fluid given by equation (\ref{eq:emtensor}), it would not be out of place to appreciate its derivation. The standard way to derive the tensor (\ref{eq:emtensor}) is the following. First, it is defined in the absence of gravity, i.e., in special relativity (SR). Then, the 
general expression for the tensor in the presence of gravity, in coordinates, say $x^\alpha$ (in which the metric tensor is identified as $g_{\mu\nu}$), is imported from SR through a coordinate transformation. The bridge between the ideal case of SR and the actual case in the presence of gravity, is provided by a locally inertial coordinate system (LICS) which can always be found at any point of spacetime (thanks to the principle of equivalence).
Let us consider an LICS $x^\alpha_\star$ which, by definition, provides an infinitesimal spacetime neighbourhood around the selected point wherein the metric tensor $g_{\mu\nu}$ does not change significantly (and reduces to the Lorentzian metric $\eta_{\mu\nu}$). Hence $\partial g_{\mu\nu}/\partial x^\alpha_\star=0$ and the effects of gravity disappear in the neighbourhood (but not in a finite region). 
This LICS is so chosen that a perfect fluid element is at rest in the neighbourhood of the position and time of interest. As an observer at the selected point sees the fluid around him isotropic, the energy density-stress tensor $T^{\mu\nu}_\star$ (in the coordinates $x^\alpha_\star$) takes the form characteristic of spherical symmetry:
\be
T^{00}_\star=\rho, ~~T^{11}_\star=T^{22}_\star=T^{33}_\star=p ~~\text{and}~~ T^{\mu\nu}_\star=0 ~~\text{for} ~~\mu\ne\nu,
\ee 
where $\rho$ and $p$ are, respectively, the energy density and pressure of the fluid measured by the considered observer. As the fluid element is at rest in the above-mentioned neighbourhood, the spatial components of the 4-velocity vector then vanish, i.e.
\be
u_x\equiv\frac{dx}{d\tau}=0, ~u_y\equiv\frac{dy}{d\tau}=0, ~u_z\equiv\frac{dz}{d\tau}=0 ~ ~{\rm and}~~
\frac{dt}{d\tau}=1,\label{eq:vel}
\ee
where we have specified $x^0_\star\equiv ct$, $x^1_\star\equiv x$, $x^2_\star\equiv y$, $x^3_\star\equiv z$; and the proper time interval $d\tau=ds/c$.
In the general coordinate system $x^\alpha$, the energy density-stress tensor $T^{\mu\nu}$ is then given by 
\be
T^{\mu\nu}=\frac{\partial x^\mu}{\partial x^\alpha_\star} \frac{\partial x^\nu}{\partial x^\beta_\star} T^{\alpha\beta}_\star, 
\ee
which reduces to (\ref{eq:emtensor}) by noting that 
 $g^{\mu\nu}=\eta^{\alpha\beta} (\partial x^\mu/\partial x^\alpha_\star)(\partial x^\nu/\partial x^\beta_\star)$
and $dx^\mu/ds=\partial x^\mu/\partial x^0_\star$, owing to the relations in (\ref{eq:vel}).
Equation (\ref{eq:emtensor}) thus derived, is valid at all points, as an LICS can always be found at any point of spacetime so that the fluid element is at rest in an infinitesimal spacetime neighbourhood around the point.

After establishing the general expression for the energy density-stress tensor of the perfect fluid, let us study its divergence which is famous for describing the mechanical behaviour of the fluid.
As our main motive is to understand the mysterious gravitational effects of the pressure of the fluid without implicating the gravitational energy, let us consider the same LICS which has been used to derive (\ref{eq:emtensor}) so that the subtleties of gravitation and the gravitational energy disappear locally in this coordinate system. Let us reconsider the same fluid element, which is at rest in the chosen spacetime neighbourhood. Though the spatial components of the 4-velocity vector vanish in this neighbourhood, the derivatives of the velocity will not be zero in general, except for its temporal component, which will vanish, as we see in the following:
the general formula for the interval,~ $ds^2=g_{\mu\nu}~dx^\mu_\star~dx^\nu_\star$ implies that 
\[
g_{00}\left(c\frac{dt}{ds}\right)^2+g_{11}\left(\frac{dx}{ds}\right)^2+....
+2g_{01}c\frac{dt}{ds}\frac{dx}{ds}+....
\]
\[
+g_{33}\left(\frac{dz}{ds}\right)^2=1,
\]
which on differentiation gives
\be
\frac{\partial}{\partial x^\alpha_\star}\left(\frac{dt}{ds}\right)=0,\label{eq:velt}
\ee
by virtue of the relations in  (\ref{eq:vel}) and by recalling that
$\partial g_{\mu\nu}/\partial x^\alpha_\star=0$, $g_{\mu\nu}=\eta_{\mu\nu}$ in the chosen neighbourhood. 
Now we assume a vanishing divergence of (\ref{eq:emtensor}) 
\be
T^{\mu\nu}_{~ ~;\nu}=0,
\ee
which, in the chosen coordinates, is given by
\be
\frac{\partial T^{\mu\nu}}{\partial x^\nu_\star}=0.\label{eq:consf}
\ee
The temporal component of equation (\ref{eq:consf}) for the case $\mu=0$, can be written in accordance with (\ref{eq:vel}) and (\ref{eq:velt}) as
\be
\frac{d}{dt}(\rho \delta v)+p\frac{d}{dt}(\delta v)=0,\label{eq:mu0}
\ee
where $\delta v=\delta x\delta y\delta z$ is the proper volume of the fluid element. The usual interpretation of this equation is that the rate of change in the energy of the fluid element is given in terms of the work done against the external pressure. This seems reasonable at the first sight, but cracks appear in it after a little reflection. The first concern, as also noticed by Tolman \cite{tolman}, is that the fluid of a finite system can be divided into similar fluid elements and the same equation (\ref{eq:mu0}) can be applied to each of these elements, meaning that proper energy $(\rho \delta v)$ of every element is decreasing when the fluid is expanding or increasing when the fluid is contracting. This leads to a paradoxical result that the sum of the proper energies of the fluid elements which make up an isolated system, is not constant! Tolman overlooked this problem by assuming a possible role of the gravitational energy in it. We however remind that there cannot be any role of the gravitational energy in equation (\ref{eq:mu0}) which has been derived in an LICS.

Another more important issue related with equation (\ref{eq:mu0}) is the following. We have already mentioned that the term $\rho$ includes in it all the possible sources of mass and energy (excluding the gravitational energy). Hence, so is included in it the energy equivalent to the work done against the external pressure. If we already know that energy (in the form of work done) is being supplied to the system or getting released from it (which can be calculated by checking by what amount $\delta v$ is increasing or decreasing), why can't this too be taken care of by the term $\rho$? There is no natural law which dictates that $\rho $ cannot include particular types of energies. So, if $\rho $ includes the energy equivalent to the work done against the external pressure, the additional work contained in the second term of equation (\ref{eq:mu0}) violates the conservation of energy!

\bigskip
Yet another paradox has been discovered recently \cite{vishwa} 
which demonstrates, more vividly, subtle inconsistencies in (\ref{eq:emtensor}). This can be derived for the spatial components of equation (\ref{eq:consf}).
By the use of (\ref{eq:vel}) and (\ref{eq:velt}) in (\ref{eq:consf}), we can write for the case $\mu=1$, the following equation
\be
\frac{\partial p}{\partial x}+\frac{\left(\rho+p\right)}{c^2}\frac{du_x}{dt}=0,
\label{eq:mu1}
\ee
which has also been obtained by Tolman in some other context.
Here $du_x/dt=du_x/d\tau=\partial u_x/\partial t$ (in the chosen coordinates)
is the acceleration of the fluid element in the x-direction. 
As any role of gravity is absent in this equation, it can be interpreted as the relativistic analogue of the Newtonian law of motion: the fluid element of unit volume, which moves under the action of the force $\partial p/\partial x$, has got the inertial mass $(\rho+p)/c^2$. But it is surprising that
 the inertial mass of the fluid element has got an additional contribution from $p$, though 
without any apparent source! Equation  (\ref{eq:mu1}), taken at the face value, reveals that $p$ should be carrying some kind of energy (density) as $p/c^2$ contributes to the inertial mass (density) of the fluid element.
However, as the term $\rho$ includes in it all the sources of energy-mass, so if $p$ `somehow' contains energy, it must be at the cost of violating the celebrated law of the conservation of energy! Though equation (\ref{eq:mu1}) is not an energy conservation
equation, but that does not allow it to defy the law of conservation of energy.

\bigskip
Finally, we derive the Tolman integral (\ref{eq:rmass}) in an LICS wherein it reduces to
\be
E=mc^2=\int (T_0^0-T_1^1-T_2^2-T_3^3)~d^3x=\int (\rho +3p)~dV,\label{eq:rmassin} 
\ee
which does not contain any gravitational energy, but still pronounces the Tolman paradox,
though the formula may be valid for a sufficiently small volume of the fluid. From equations\footnote{Equation (\ref{eq:mu0}) can be written alternatively as
$\delta v ~d\rho/dt+(\rho+p)d(\delta v)/dt=0.$} (\ref{eq:mu0}), (\ref{eq:mu1}) and (\ref{eq:rmassin}), we notice
something unexpected: we still encounter different (unequal) mass densities
$(\rho +3p)/c^2$ and $(\rho +p)/c^2$ of the same fluid element though there remains no possibility, in the present situation, of any contribution from the gravitational energy to these masses to make them equal, as they are derived in an LICS!

Although the equality of the  inertial and gravitational masses is the starting point of any metric theory of gravitation, however, the resulting theories have not achieved their own defining feature unequivocally. The two gravitational masses, the active and the passive ones, have remained unequal and the equality of the inertial and the active gravitational masses have remained a dream.
The usual interpretation thereof seeks refuge in the gravitational energy.
However, the present analysis ensures that there is no possibility of any role of the gravitational energy
in making the different masses equal. Perhaps the origin of this problem is elsewhere.

It is thus established that the relativistic description of matter given by (\ref{eq:emtensor}) suffers from some subtle inherent inconsistencies in its basic formulation. The point to note is that there is no role of the notorious (pseudo) energy of the gravitational field in these problems.

\section{Corrections and Their Consequences}

There already exists the resolution of the Tolman paradox. Let us take tips from there to resolve the other paradoxes described above and study the consequences.
The standard resolution of the Tolman paradox, first presented by  Misner \& Putnam \cite{misner} in 1959,
is provided through the following prescription: ``If the pressure is confined by non-gravitational means, there must be tension (negative pressure) in the walls (to keep the fluid inside and the field static) which will compensate the additional energy causing the paradox.'' In fact, the integrated value of the negative pressure ($-p=-T_i^i$, $i=1, 2, 3$, no summation over i) will contribute to (\ref{eq:rmass}) a term which just counterbalances the increased mass arising from the $3p=(T_1^1+T_2^2+T_3^3)$ term. More recently this resolution has been confirmed by Ehlers et al. \cite{ehlers2} in a more refined version by giving a tensorial treatment to the surface tension of the walls and the trapped fluid sphere.

It should be noted that the key point of these resolutions $-$ the counterbalancing the pressure of the fluid by stresses in the wall $-$ does not emanate either from the derivation of $T^{\mu\nu}$ or from the underlying gravitational theory. Rather it has been {\it put by hand}  as a correction. After getting corrected, equation (\ref{eq:rmass}) reduces to the classical result
\[
~~~~~~~~~~~~~~~~~~~~~~~~~~~~~~~~~~~~~~~
E=\int \mathcal{T}_0^0~d^3x= \int \rho\sqrt{|g_{00}|}~dV.
\qquad ~~~~~~~~~~~~~~~~~~~~~~~~~~~~~\text{(4$^\prime$)}
\]

Let us apply this clue to the paradoxes represented by equations (\ref{eq:mu0}) and (\ref{eq:mu1}), for the case when the system is held by non-gravitational means\footnote{The nature of the corrections is not clear when the system is held by gravitation. 
Should one consider the systems held by the gravitation and those by non-gravitational means, equivalent? Even in the situation when the system is held by gravitation, one would not get any contribution from the gravitational energy in equations (\ref{eq:mu0})$-$(\ref{eq:rmassin}) (as they are derived in an LICS), but still sufferring from the paradoxes. This shows the equivalence of the cases held by gravitation and non-gravitational means, so far as the paradoxes are concerned.} .
 As the tension of the wall is transmitted to the fluid element, each of the $T_x^x$, $T_y^y$ and $T_z^z$ present in the considered fluid element, is balanced by a -$T_i^i$ from the wall.
Hence the term $pd(\delta v)/dt$ in equation (\ref{eq:mu0}) [which can also be written as $pd(\delta v)/dt=\delta y\delta z~T_x^xd(\delta x)/dt+\delta z\delta x~T_y^yd(\delta y)/dt+\delta x\delta y~T_z^zd(\delta z)/dt$] is balanced by $-\delta y\delta z~T_x^xd(\delta x)/dt-\delta z\delta x~T_y^yd(\delta y)/dt-\delta x\delta y~T_z^zd(\delta z)/dt=-pd(\delta v)/dt$, modifying equation (\ref{eq:mu0}) to
\[
~~~~~~~~~~~~~~~~~~~~~~~~~~~~~~~~~~~~~~~~~~~~~~~~~~~
\frac{d}{dt}(\rho \delta v)=0,
\qquad ~~~~~~~~~~~~~~~~~~~~~~~~~~~~~~~~~~~~~~~~\text{(11$^\prime$)}
\]
which is now free from the paradoxes. We notice something very remarkable in these corrections. The corrections made in equations (\ref{eq:rmass}) and (\ref{eq:mu0}) are effectively equivalent to replacing the active gravitational mass density $(\rho+3p)/c^2$ as well as the inertial mass density $(\rho+p)/c^2$ (which is also the passive gravitational mass density) by the simple (total) Newtonian mass density
$\rho/c^2$. {\it Hence the long-sought-after validity of the principle of equivalence is achieved naturally here!} This modifies equation (\ref{eq:mu1}) to
\[
~~~~~~~~~~~~~~~~~~~~~~~~~~~~~~~~~~~~~~~~~~~~~~~~~
\frac{\partial p}{\partial x}+\frac{\rho}{c^2}\frac{du_x}{dt}=0,
\qquad ~~~~~~~~~~~~~~~~~~~~~~~~~~~~~~~~~~~~~~\text{(12$^\prime$)}
\]
which is equivalent to balancing the term $(p/c^2)(du_x/dt)$ of equation (\ref{eq:mu1}) by $-(p/c^2)(du_x/dt)$.
We notice that there is only one contribution from $T_i^i$s, viz. from $T_x^x$, to equation  (\ref{eq:mu1}) whereas all the three $T_i^i$s contribute to equations (\ref{eq:rmass}) and (\ref{eq:mu0}). The reason is that equation (\ref{eq:mu1}) represents a balance of forces along the x-direction only. Similar contributions from $T_y^y$ and $T_z^z$ would appear to the other two similar equations  derived from equation (\ref{eq:consf}) for the cases $\mu=2$ and $\mu=3$ respectively. Obviously, equation (\ref{eq:rmassin}) gets corrected to
\[
~~~~~~~~~~~~~~~~~~~~~~~~~~~~~~~~~~~~~~~~~~~~~
E=\int T_0^0~d^3x= \int \rho ~dV.
\qquad ~~~~~~~~~~~~~~~~~~~~~~~~~~~~~~~~~\text{(13$^\prime$)}
\]

The corrected equations are  convincing and free from the conceptual problems. Additionally, they ascertain the equality of the gravitational and inertial masses, which is remarkable. However, replacing the different general-relativistic energy densities, appearing in the equations of gravitation and cosmology, by the total energy density $\rho$, would be revolutionary.

\subsection{Any Observational Support for $T^{\mu\nu}$?}

Does it then mean that Einstein's `wood' is not only low grade compared to
the standards of his `marble' but it is also rotten and infected? 
 It should be noted that the relativistic description of the matter, in the form 
of the energy density-stress tensor (\ref{eq:emtensor}), has never been tested in any direct experiment. The classical tests of GR consider $\rho = p = 0$. The
same is true for the more precise tests of GR made through the observations
of the radio pulsars, which are rapidly rotating strongly magnetized neutron
stars. The pulsar tests assume the neutron stars as point-like objects and
look for the relativistic corrections in the post-Keplerian parameters by
measuring the pulsar timing. The test does not even require to know the
exact nature of the matter that pulsars and other neutron stars are made
of.
Hence, the celebrated tests of GR have remained limited to test only the geometric aspect of the theory.

\bigskip
Although the above-discussed theoretical crisis in the relativistic
formulation of matter is too striking to be ignored, one should be
cautious in pronouncing a judgment against a theory like GR, that has been
agreed upon by experts for a
century and is regarded a highly successful theory of gravitation
in terms of its agreement with experimental results and its ability to
predict new phenomena. Although the experimental tests, which have
verified the theory of GR, have been carried out in empty
space, however
the following two points may be considered in support of the energy density-stress tensor of the
perfect fluid.

\begin{quote}
\begin{enumerate}
\item
The general-relativistic formulation of the disordered radiation (which is a
particular case of the energy density-stress tensor of the perfect fluid) is
consistent with its formulation resulting from
Maxwell's theory of classical electrodynamics.

\item
Although the standard interpretation of the observations of the bending of
starlight, when it passes past the Sun, is given in terms of the correct
geometry resulting from the Einstein vacuum field equations $R^
{\mu\nu}=0$ (Schwarzschild solution), however, the observed deflection, twice as much as predicted by a heuristic argument made in
Newtonian gravity, appears roughly consistent with Tolman's formula (\ref{eq:rmass}) for the disordered radiation \cite{tolman}\footnote{Ironically a result derived by Tolman himself (page 250 of \cite{tolman}) appears contradictory to this. In a weak field, like that of the sun, where Newtonian gravitation can be regarded as a satisfactory approximation, equation (\ref{eq:rmass}) can be written as
$E=\int \rho dV +(1/2c^2)\int \rho \psi dV,$
where $\psi$ is the Newtonian gravitational potential. As $\psi$ is negative, we note that the general relativistic active gravitational mass $E/c^2$ of the gravitating body is obviously less than its proper mass $(1/c^2)\int \rho dV$ and is expected to give a lower value for the gravitational deflection of light!}.  

\end{enumerate}
\end{quote}

\section{Conclusion}

A critical analysis of the relativistic formulation of matter in terms of $T^{\mu\nu},$ reveals some
surprising inconsistencies and paradoxes. Corrections are discovered which
lead to the long-sought-after equality of the gravitational and inertial
masses, which are otherwise different in the metric theories of gravitation.
Although these obligatory corrections (which must be performed in order to rectify the relativistic formulation of matter) are important in their own right, they do not seem to be consistent with many cosmological observations. For example,  they render the standard cosmology altogether incompatible with an accelerating universe and hence with the supernovae Ia observations. This indicates that perhaps the relativistic description of matter by the $T^{\mu\nu}$ is not compatible with the geometric description of the metric theories \cite{submitted}.

We have tried to provide an overview of the situation resulting from the
newly discovered paradoxes, posing serious concern for the modern theories of gravitation, amid the evidences
which defend them.
It seems likely that an entirely new paradigm is required to
resolve this crisis which may completely revolutionize our understanding of
the theory.
The resolution will certainly require additional ideas and a critical re-examination of the many simplifying
assumptions underlying present scenarios.

\bigskip

\section{Appendix}
It is only unto the experimental tests to decide the final validity of a theory. 
While scientists are crafting experiments to measure already tested predictions of GR with ever-greater precision,
it is surprising that not a single experiment has been devised so far to test the gravitational effect of pressure.
One such test is also warranted by the present crisis in the standard cosmology, resulting from the discovery of the  mysterious `dark energy' (conjured up to explain the current observations), which poses a serious confrontation between fundamental physics and cosmology. It may be mentioned that
the most exotic property of the dark energy is its negative pressure, which is ultimately related with the gravitational effect of pressure. In the following, we propose a simple experiment to test this novel prediction of the geometric theories of gravitation. 

\bigskip
\noindent
{\bf An experiment to test the gravitational effect of pressure:}\\
In order to test directly if pressure does carry energy density and hence makes gravitational contribution, we propose a simple Cavendish-type experiment in which the attractor masses, of the torsion balance, are replaced with hollow spherical containers containing some source material (say, $\cal{S}$) which can be converted into gas when required. The experiment is to be performed in two steps. In the first step, one performs the experiment with $\cal{S}$ in its solid or liquid state (when its pressure is zero). In the second step, one repeats the experiment with $\cal{S}$ converted into gas (which has a non-zero pressure). If pressure does carry energy and hence makes a gravitational contribution, we should expect a difference between the readings of the two steps.  Of course, in either case, one would need to achieve a perfect absence of any heat transfer between the containers and the surroundings. 
The recommendations for the material $\cal{S}$ are (i) a material that sublimes from solid
 to gas, such as solid carbon dioxide; or (ii) two
 liquids or a liquid plus a solid that would react to produce a gas, analogous to the expansion of an automobile air bag. For a volume $V$ of the container, we expect a gravitational contribution equivalent to an effective mass $pV/c^2$ from the pressure $p$ of the gas.

We can estimate this contribution by approximating the gas with an ideal gas. If we consider $x$ kg of $\cal{S}$ and produce a gas of molecular mass $M$, the gas would contain $n=xN/M \times10^3$ molecules, where $N\approx 6\times10^{23}$ is Avogadro's number. Hence  we should expect an energy contribution of
$pV=nk_BT\approx 25x/M\times10^5$ J at the room temperature ($k_B\approx1.38\times10^{-23}$ JK$^{-1}$ is the Boltzman constant) and hence a mass contribution of $pV/c^2\approx 2.8 x/M\times 10^{-11}$kg. Thus the fractional contribution to the mass of  $\cal{S}$  from the pressure of the gas $\approx 2.8/M\times 10^{-11}$. For a gas, like carbon dioxide ($M=44$), this comes out as $\approx 10^{-12}$ (which is perhaps too small to be measured by the present means).

This feeble effect can be significantly enhanced by heating the gas (say, by some electrical appliance fitted with the containers). Suppose we supply $y$ J of heat to the gas, this is imparted to the kinetic energy density of its molecules. As the magnitude of the pressure of an ideal gas $=\frac{2}{3}\times {\rm its ~kinetic ~energy ~density}$, the fractional increase in the mass of the gas, from this effect, would be roughly
\[
~~~~~~~~~~~~~~~~~~~~~~~~~~~~~~~~~~~~~~~~~~~~~~
\frac{x\times10^{-12}+0.67y/c^2}{x+y/c^2}, 
\qquad ~~~~~~~~~~~~~~~~~~~~~~~~~~~~~~~~\text{(A.1)}
\]
where $c$, the speed of light, is measured in meter per second. Hence, it is clear that if we can increase $y$ significantly (higher than $10^5$ J), it would not be as difficult to detect (A.1) as is the meagre contribution of the order $10^{-12}$. This will test not only the gravitational effect of pressure but also the gravitational affirmation of the mass-energy equivalence which has remained just an extrapolation from special relativity to GR and needs to be tested.

\vspace{1cm}
\noindent
{\bf Acknowledgement:} The author thanks Vijay Rai for making available some
important literature.


\noindent
\begin{references}





\bibitem{MTW} C. W. Misner, K. S. Thorn and J. A. Wheeler, {\it Gravitation} (W. H. Freeman and Company, New York, 1970).

\bibitem{PhysRep} S. Capozziello and M. De Laurentis, Phys. Rep. {\bf 509} (2011) 167.

\bibitem{Ehlers} J. Ehlers, I. Ozsvath and E. L. Schucking, Am. J. Phys. {\bf 74} (2006) 607 (arXiv: gr-qc/0505040). 

\bibitem{tolman}  R. C. Tolman, {\it Relativity, Thermodynamics and
Cosmology} (Oxford University Press, 1934).

\bibitem{harrison}  E. R.Harrison, Astrophys. J. {\bf 446}, 63, 1995.

\bibitem{vishwa} R. G. Vishwakarma, Astrophys. Space Sci. {\bf 321} (2009) 151 (arXiv:0705.0825).

\bibitem{misner} C. W. Misner and P. Putnam, Phys. Rev. {\bf 116} (1959) 1045.

\bibitem{ehlers2}  J. Ehlers, I. Ozsvath, E. L. Schucking and Y. Shang, Phys. Rev. D {\bf 72} (2005) 124003 (arXiv: gr-qc/0510041).

\bibitem{submitted} R. G. Vishwakarma, {\it ``Mysteries of the Geometrization of Gravitation"} (submitted).


\end{references}
\end{document}